\documentclass[aps,prl,showpacs,twocolumn,nofootinbib,superscriptaddress,amsmath,amssymb]{revtex4}

\usepackage{verbatim} 
\usepackage{amsmath}
\usepackage{amssymb}
\usepackage{braket}

\usepackage{color}

\newcommand{\be}{\begin{eqnarray}}
\newcommand{\ee}{\end{eqnarray}}

\begin{document}

\title{Chiral torsional effect}

\author{Z.\,V.\,Khaidukov}
\affiliation{Institute for Theoretical and Experimental Physics, B. Cheremushkinskaya 25, Moscow, 117259, Russia}
\affiliation{Moscow Institute of Physics and Technology,
Institutskiy per. 9, 141700 Dolgoprudny, Moscow Region, Russia}

\author{M.A. Zubkov \footnote{On leave of absence from Institute for Theoretical and Experimental Physics, B. Cheremushkinskaya 25, Moscow, 117259, Russia}}
\email{zubkov@itep.ru}
\affiliation{Physics Department, Ariel University, Ariel 40700, Israel}

\date{\today}

\begin{abstract}
We propose the new nondissipative transport effect - the appearance of axial current of thermal quasiparticles in the presence of background gravity with torsion. For the non-interacting model of massless Dirac fermions the response of the axial current to torsion is derived. The chiral vortical effect appears to be the particular case of the chiral torsional effect. The proposed effect may be observed in the condensed matter systems with emergent relativistic invariance (Weyl/Dirac semimetals and the $^3$He-A superfluid).  
\end{abstract}

\maketitle

\section{Introduction}

The non - dissipative transport effects have been widely discussed recent years \cite{Landsteiner:2012kd,semimetal_effects7,Gorbar:2015wya,Miransky:2015ava,Valgushev:2015pjn,Buividovich:2015ara,Buividovich:2014dha,Buividovich:2013hza}. These effects are to be observed in the non - central heavy ion collisions~\cite{ref:HIC}. They have also been considered for the  Dirac and Weyl semimetals \cite{semimetal_effects6,semimetal_effects10,semimetal_effects11,semimetal_effects12,semimetal_effects13,Zyuzin:2012tv,tewary,16} and in $^3$He-A \cite{Volovik2003}.
It was expected that among the other effects their family includes  the chiral separation effect (CSE) \cite{Metl}, the chiral magnetic effect (CME)  \cite{CME,Kharzeev:2015znc,Kharzeev:2009mf,Kharzeev:2013ffa}, the chiral vortical effect (CVE) \cite{Vilenkin}, the anomalous quantum Hall effect (AQHE)  \cite{TTKN,Hall3DTI,Zyuzin:2012tv}. All those phenomena have the same origin - the chiral anomaly. The naive "derivation" of the CME from the standard expression for the chiral anomaly has been presented, for example, in \cite{Zyuzin:2012tv,CME,SonYamamoto2012}.

However, the more accurate consideration demonstrates that the original equilibrium version of the CME does not exist. In \cite{Valgushev:2015pjn,Buividovich:2015ara,Buividovich:2014dha,Buividovich:2013hza} this subject has been elaborated via the numerical lattice methods. In the condensed matter theory the absence of CME was reported in \cite{nogo}. The same conclusion has been obtained on the basis of the no - go Bloch theorem \cite{nogo2}. The analytical proof of the absence of the equilibrium CME was presented in \cite{Z2016_1,Z2016_2} using the technique of Wigner transformation \cite{Wigner,star,Weyl,berezin}  applied to the lattice models\footnote{The CME may, however, survive in a certain form as a non - equilibrium kinetic phenomenon - see, for example,  \cite{ZrTe5}.}. The same technique allows to reproduce the known results on the AQHE \cite{Z2016_1}. It also confirms the existence of the equilibrium CSE \cite{KZ2017}.
Notice, that in the framework of the naive nonregularized quantum field theory the CSE was discussed, for example, in \cite{Gorbar:2015wya} in the technique similar to the one that was used for the consideration of the CME \cite{CME,Zyuzin:2012tv}.

The investigation of momentum space topology was developed previously mainly within the condensed matter physics. It allows to relate the gapless boundary fermions to the bulk topology for the topological insulators and to describe the stability of the Fermi surfaces of various kinds. Besides, this technique allows to calculate the number of the  fermion zero modes on vortices (for the review see \cite{Volovik2003,Volovik:2011kg}). Momentum space topology of QCD has been discussed recently in \cite{Z2016_3}. The whole Standard Model of fundamental interaction was discussed in this framework in \cite{Volovik:2016mre}.

Recently the new non - dissipative transport effects were proposed:  the rotational Hall effect \cite{RotHall} and the scale magnetic effect \cite{SME}. In the present paper we will propose one more non - dissipative transport effect - the chiral torsional effect. Namely, we will discuss the emergence of  axial  current in the presence of torsion. It will be shown that this effect is intimately related to the chiral vortical effect \cite{Vilenkin}. In conventional general relativity \cite{vlad} torsion vanishes identically, it appears only in its various extensions. The background (non - dynamical) gravity with torsion emerges, in particular,  in certain condensed matter systems.  For example, elastic deformations in graphene and in Weyl semimetals induce the effective torsion experienced by  the quasiparticles \cite{VolZub,Voz}. In $^3$He-A torsion appears dynamically when motion of the superfluid component is non - homogeneous. Below we will consider mostly the massless free Dirac fermions in the presence of the background gravity with torsion. We assume that this model is the  relevant ingredient of the low energy effective field theory for the appropriate condensed matter system (either the Weyl/Dirac semimetal or $^3$He-A) or of the theory of elementary particles in the presence of real background gravity with torsion. The other ingredients of the mentioned theories are interactions between the fermions due to the exchange by various interaction carriers including the gauge bosons. On the present level of consideration we omit the interactions completely.

Besides, we disregard the natural anisotropy existing without emergent gravity in $^3$He-A and Weyl semimetals. Therefore, while applying expression for the chiral torsional effect obtained below to these materials one should take care of this anisotropy. It actually may be restored in the final expression in the straightforward way.

\section{Response of the axial current to torsion}

We are considering the model of massless Dirac fermions, which serves as the low energy approximation for the electronic excitations in Weyl semimetals and also for the fermionic excitations in $^3$He-A. For simplicity we will restrict ourselves by one flavor of Dirac fermions, and do not take into account the anisotropy of the Fermi velocity. In order to take it into account we should simply rescale the coordinates in an appropriate way.

Following  \cite{Wigner,CSEwith} we will calculate here the response of the axial current $j^{5\mu}(x)=\bar{\psi}(x)\gamma^{\mu}\gamma^{5}\psi(x)$  to torsion in terms of the Wigner transformation of the Green function. We consider the case of vanishing spin connection and the nontrivial torsion encoded in the vielbein $e^a_\mu(x) = \delta^{a}_{\mu}+\delta e^{a}_{\mu}(x)$. The inverse vielbein is denoted here by $E_a^\mu = \delta_a^\mu + \delta E_a^\mu$, it obeys equation $ e^a_\mu(x)  E_b^\mu(x) = \delta_b^a $, which gives $\delta e^a_b(x)  \approx -\delta E^a_b  $. Momentum operator is to be substituted by $p_{a} \to (\delta_{a}^{b}+\delta E^{b}_{a}(x))p_{b} \approx (p_{a}-\delta e^{b}_{a}(x)p_{b} )$. Product of the coordinate - depending vielbein and momentum operator should be symmetrized in order to lead to the Hermitian Hamiltonian  of the quasipaticles \footnote{In this form gravity emerges in graphene and Weyl semimetals.}.  The term $\delta e^{b}_{i}p_{b}=A_{i}$  to a certain extent is similar to the gauge field (see also  \cite{Sumiyoshi:2015eda,ACVE}). Suppose, that $G(p)$ is the two point fermion Green function in momentum space for the system of Dirac spinors. In order to obtain expressions for the axial  current we deform it as follows:
\be G(p)\to G(p-B(x)\gamma^{5})
\ee
Next, the axial current is to be expressed as the functional derivative of the partition function with respect to $B$. The partition function is built using the Dirac operator defined in momentum space as $G^{-1}(p-B(x)\gamma^{5})$. In this consideration $x$ is to be treated as parameter.  Following the same steps as those of \cite{CSEwith} in the leading order in the derivative of the vielbein we   obtain the following expression for the axial current:
\be j^{5k}&=&\frac{i}{2}T^{a}_{ij} \int{\frac{d\omega d^3p}{(2\pi)^{4}}p_{a}{\rm Tr}\, G\partial_{p_{i}}G^{-1}\partial_{p_{j}}G\gamma^{5}\partial_{p_{k}}G^{-1}} \label{mainform}\\
 &&T^{a}_{ij}=\partial_{i}e^{a}_{j}-\partial_{j}e^{a}_{i} \nonumber
\ee
Let us consider the situation when the only nonzero components of torsion are given by:
\be T^{0}_{ij} = \epsilon^0_{.fij} S^{f}
\ee
where $\bf S$ is a certain three - vector.
 In   case of the non-interacting theory of massless Dirac fermions in continuous space in the presence of finite temperature $T$ we obtain the following expression:
\be j^{5k}=4 S^{k}T\sum_{\omega_{n}}\int{\frac{d^{3}p}{(2\pi)^3}\frac{\omega^{2}_{n}}{(\omega^{2}_{n}+p^{2})^2}} \label{main}
\ee
Here the sum is over the discrete Matsubara frequencies $\omega_n = 2\pi (n+1/2)/T$, $n\in Z$.
 In the expression of Eq. (\ref{mainform}) we  do not see the signature of the topological protection from  various corrections. Therefore, we do not exclude that the small deformations of the Green's function may lead to the variation of the axial current.  Thus, there will, possibly, appear  corrections to this expression due to interactions.

\section{Sum over the Matsubara frequencies}

The obtained above expression (\ref{main}) is divergent, and requires regularization.
Below we present the results of our calculations made using the two complementary methods: via the  zeta-regularization and via the sum over the Matsubara frequencies with the contribution of vacuum subtracted. In both cases the results coincide. Let us consider first the direct summation over the Matsubara frequencies:
    \be j^{5z}=4 S^{3}T\sum_{\omega_{n}}\int{\frac{d^{3}p}{(2\pi)^3}\frac{\omega^{2}_{n}}{(\omega^{2}_{n}+p^{2})^2}}=j^{5z}_{vac}+j^{5z}_{medium} \label{eqmats} \ee
where $ \omega_{n}=2\pi T(n+1/2)$. Let us perform summation over $n$:
\begin{eqnarray}
&&4 T \sum_{\omega_{n}}\frac{\omega^{2}_{n}}{(\omega^{2}_{n}+p^{2})^2} = - 2 \int \frac{dz}{2\pi i}\frac{z^2}{(z^2-p^2)^2}{\rm th}\Big(\frac{z}{2T}\Big) \nonumber\\ &&=\frac{2z^2}{(z+p)^2}\frac{d}{dz}{\rm th}\Big(\frac{z}{2T}\Big)\Big|_{z=p}+\frac{2z^2}{(z-p)^2}\frac{d}{dz}{\rm th}\Big(\frac{z}{2T}\Big)\Big|_{z=-p} \nonumber\\&&\quad + \frac{1}{p}{\rm th}\Big(\frac{p}{2T}\Big)\nonumber\\&&=  \frac{1}{2T {\rm ch}^2 \Big(\frac{p}{2T}\Big)} + \frac{1}{p}{\rm th}\Big(\frac{p}{2T}\Big)
\end{eqnarray}
In the first row the integral is over the contour surrounding all poles of the hyperbolic tangent. This contour is then deformed to surround points $\pm p$.
In this expression we  separate  the contribution of thermal quasiparticles subtracting the contribution of vacuum (the expression of Eq. (\ref{eqmats}) for $T=0$). We denote $x=\frac{p}{2T}$ and obtain:
		\be j^{5i}_{medium}=2T^{2}} S^{i}\int_0^\infty{\frac{ x^{2}dx}{\pi^2} \Big( \frac{1}{ {\rm ch}^2 x} +  \frac{1}{x}{\rm th} x - \frac{1}{x}\Big)\ee
Vacuum part is formally divergent, and we do not consider it here.
		After integration over $x$  we obtain the following expression for the separate contribution of medium (i.e. of the thermal quasiparticles):
		\be
		{j}^{5k}_{medium}=\frac{T^{2}}{{12}}{S}^k = -\frac{T^{2}}{{24}}\epsilon^{0kij}T^0_{ij}\label{CTE}
		\ee

It is well - known that $\zeta$  - regularization subtracts systematically all divergencies. It allows to include all logarithmic divergencies to the renormalization of coupling constants automatically. At the same time the power - like divergencies are subtracted from all expressions completely.
Formally integration over momenta in Eq. (\ref{eqmats}) gives $\vec{j}^{5}=\frac{1}{2\pi}\vec{S} T\sum^{\infty}_{n=-\infty}\,|\omega_n|$.
In zeta - regularization instead of this expression we obtain
$$
\vec{j}^{5}=\frac{1}{\pi}\vec{S} T\sum^{\infty}_{n=0}\,\frac{1}{\omega^s_n}\Big|_{s=-1}
$$
 Here the sum over $n$  determines the function of $s>1$.  The final answer is given by its analytical continuation  to $s=-1$. This results in the following resummation
 		\be \vec{j}^{5}&=&2\vec{S} T^{2}\Big(\sum_{n=0}^{\infty}\frac{1}{(n+1/2)^{s}} \Big)_{s=-1}\nonumber \\&=&
2\vec{S} T^{2}\zeta_H(-1,1/2) =\frac{T^{2}}{12}\vec{S} \label{eqdzeta}
		\ee
 In the second row $\zeta_H(s,\nu)$ is the Hurwitz zeta - function.

 One can see that the expressions for the current produced by the thermal quasiparticles of (\ref{eqmats})  and (\ref{eqdzeta}) coincide.

	\section{CVE effect as the particular case of the chiral torsional effect}

Let us consider the particular non - relativistic system (having the emergent relativistic symmetry at low energies) in the presence of macroscopic motion with velocity $\bf v$. When $\bf v$ is constant or slowly varying, the one - particle Hamiltonian is given by $\hat{H} = \hat{H}_0 + {\bf p} {\bf v}$. Here $\hat{H}_0$ is the Hamiltonian in the reference frame accompanying the motion.  For the case of emergent massless Dirac fermions $\hat{H}_0 = \gamma^0 \sum_{k=1,2,3}\gamma^k p_k $. Then the combination $p_0 -  \hat{H}$ may be written as $\gamma^0 E_a^\mu \gamma^a p_\mu $ with the inverse vierbein $E^0_0 = 1$, $E_a^k = \delta_a^k$, $E_0^k = -v_k$, $E^0_a=0$ for $a,k = 1,2,3$.   Therefore, we are able to treat velocity of the macroscopic motion as the components of the vielbein: $e^0_k = v_k$, $k=1,2,3$. Correspondingly, torsion appears with the nonzero components $T^0_{ij} = \partial_i v_j - \partial_j v_i  = 2\epsilon_{0ijk}\omega^k$, where $\vec{\omega}$ is vector of angular velocity.  Eq. (\ref{CTE}) gives
\be
	{j}^{5k}_{medium}= -\frac{T^{2}}{{24}}\epsilon^{0kij}T^0_{ij}=\frac{T^{2}}{{6}} \omega^k
\ee
which is nothing but the conventional expression for the chiral vortical effect
(see also \cite{Land}). Recall that the latter effect for vanishing chemical potential is the appearance of the axial current in the system of rotating thermal quasiparticles. Typically in the consideration of this effect the vacuum as a whole is assumed to be at rest, only the excitations over vacuum are rotated. Therefore, indeed, the subtraction performed above (of the vacuum contribution to the chiral torsional effect) effectively results in the description of the chiral vortical effect for vanishing $\mu$.

\section{Conclusions and discussions}

To conclude, we have calculated the axial current produced by  thermal quasiparticles in the presence of torsion for the system of the noninteracting massless Dirac fermions. Eq. (\ref{CTE}) is obtained in two ways: via zeta regularization and via the summation over Matsubara frequencies with the vacuum contribution subtracted.  Although we did not consider the effects of interactions directly, there are certain indications, that the coefficient in Eq. (\ref{CTE})  is subject to the renormalization due to interactions. Notice, that in Eq. (\ref{CTE}) there is no general coordinate invariance. First of all, this invariance is absent because we consider the system with vanishing spin connection, which corresponds to the case of emergent gravity in real Weyl materials \cite{Cortijo:2015jja}. This corresponds to the partial gauge fixing, which eliminates the general coordinate invariance.  Moreover, we considered the system with finite temperature. Therefore, even the  Lorentz invariance is broken and Eq. (\ref{CTE}) actually possesses the $O(3)$ symmetry of spacial rotations.

It is also worth mentioning that we considered only the components of torsion $T^0_{ij}$ with $i,j=1,2,3$. The possible appearance of the axial current in the presence of the other components of torsion remains open and requires an additional consideration, which is out of the scope of the present paper. The resulting current of Eq. (\ref{CTE}) represents the new anomalous transport effect. The corresponding current is non - dissipative, at least in the approximation of the slowly varying vielbein/torsion. The explanation for this is the mentioned above analogy between the composition $\delta e^0_i \omega$ and the electromagnetic vector - potential. According to this analogy ${\cal H}^k = T^0_{ij}\epsilon^{0ijk}\omega$ plays the role of the effective magnetic field. The latter does not perform the work and thus cannot lead to dissipation.

The divergency of the vacuum axial current in the presence of background torsion prompts that the quantum field theory in the presence of background gravity with torsion is ill - defined. Presumably, there are the aspects of the theory, in which it depends on regularization. In the other words, suppose, that the microscopically different real physical systems have naively the same continuum low energy effective field theory. In the presence of emergent torsion these effective theories become different. In particular, we expect the appearance of such differences between the low energy description of electrons in  Weyl semimetals (in the presence of elastic deformations) and the description  of the quasiparticles in the $^3$He-A superfluid (in the presence of the in - homogeneous motion of the superfluid component). This difference may, possibly, be related to the appearance of the Nieh - Yan - like term \cite{Nieh} in the chiral anomaly for the solids and its absence for $^3$He-A. Notice, that experiments with vortices in $^3$He-A prompt that the chiral anomaly in this superfluid does not contain the Nieh - Yan term (see \cite{Volovik2003} and references therein). At the same time the recent theoretical consideration of Weyl semimetals indicates the appearance of this term in the chiral anomaly for these materials \cite{Parrikar}.

It would be interesting to apply the lattice regularization in the spirit of \cite{KZ2017}  for the calculation of axial current of massless Dirac fermions in the presence of nontrivial vielbein. Such a calculation may shed led on various questions related to the vacuum contribution to the chiral torsional effect and to the possible appearance of the Nieh - Yan term in the chiral anomaly.

Z.V.K. is grateful to the physics department of the Ariel University for kind hospitality. The work of Z.V.K. was supported by Russian Science Foundation Grant No 16-12-10059. Both authors kindly acknowledge useful discussions with M.Suleymanov.


\begin{thebibliography}{99}



\bibitem{Landsteiner:2012kd}
  K.~Landsteiner, E.~Megias and F.~Pena-Benitez,
  ``Anomalous Transport from Kubo Formulae,''
  Lect.\ Notes Phys.\  {\bf 871} (2013) 433
  [arXiv:1207.5808 [hep-th]].

\bibitem{semimetal_effects7}
M. N. Chernodub, A. Cortijo, A. G. Grushin, K. Landsteiner, and M. A. Vozmediano,
{\sl ``A condensed matter realization of the axial magnetic effect''},
Phys. Rev. B {\bf 89}, 081407(R) (2014) [arXiv:1311.0878].


\bibitem{Gorbar:2015wya}
  E.~V.~Gorbar, V.~A.~Miransky, I.~A.~Shovkovy and P.~O.~Sukhachov,
  ``Chiral separation and chiral magnetic effects in a slab: The role of boundaries,''
  Phys.\ Rev.\ B {\bf 92} (2015) 24,  245440
  doi:10.1103/PhysRevB.92.245440
  [arXiv:1509.06769 [cond-mat.mes-hall]].

\bibitem{Miransky:2015ava}
  V.~A.~Miransky and I.~A.~Shovkovy,
  ``Quantum field theory in a magnetic field: From quantum chromodynamics to graphene and Dirac semimetals,''
  Phys.\ Rept.\  {\bf 576} (2015) 1
  [arXiv:1503.00732 [hep-ph]].



\bibitem{Valgushev:2015pjn}
  S.~N.~Valgushev, M.~Puhr and P.~V.~Buividovich,
  ``Chiral Magnetic Effect in finite-size samples of parity-breaking Weyl semimetals,''
  arXiv:1512.01405 [cond-mat.str-el].





\bibitem{Buividovich:2015ara}
  P.~V.~Buividovich, M.~Puhr and S.~N.~Valgushev,
  ``Chiral magnetic conductivity in an interacting lattice model of parity-breaking Weyl semimetal,''
  Phys.\ Rev.\ B {\bf 92} (2015) 20,  205122
  doi:10.1103/PhysRevB.92.205122
  [arXiv:1505.04582 [cond-mat.str-el]].


\bibitem{Buividovich:2014dha}
  P.~V.~Buividovich,
  ``Spontaneous chiral symmetry breaking and the Chiral Magnetic Effect for interacting Dirac fermions with chiral imbalance,''
  Phys.\ Rev.\ D {\bf 90} (2014) 125025
  doi:10.1103/PhysRevD.90.125025
  [arXiv:1408.4573 [hep-th]].



\bibitem{Buividovich:2013hza}
  P.~V.~Buividovich,
  ``Anomalous transport with overlap fermions,''
  Nucl.\ Phys.\ A {\bf 925} (2014) 218
  doi:10.1016/j.nuclphysa.2014.02.022
  [arXiv:1312.1843 [hep-lat]].

\bibitem{ref:HIC}
  L.~P.~Csernai, V.~K.~Magas and D.~J.~Wang,
  Phys.\ Rev.\ C {\bf 87}, no. 3, 034906 (2013)
  [arXiv:1302.5310 [nucl-th]];
  F.~Becattini {\it et al.},
  Eur.\ Phys.\ J.\ C {\bf 75}, no. 9, 406 (2015)
  [arXiv:1501.04468 [nucl-th]];
    Y.~Jiang, Z.~W.~Lin and J.~Liao,
  Phys.\ Rev.\ C {\bf 94}, no. 4, 044910 (2016)
  [arXiv:1602.06580 [hep-ph]];
    W.~T.~Deng and X.~G.~Huang,
  Phys.\ Rev.\ C {\bf 93}, no. 6, 064907 (2016)
  [arXiv:1603.06117 [nucl-th]].





\bibitem{semimetal_effects6}
S. Parameswaran, T. Grover, D. Abanin, D. Pesin, and A. Vishwanath,
{\sl ``Probing the chiral anomaly with nonlocal transport in Weyl semimetals},
Phys. Rev. X {\bf 4}, 031035 (2014) [arXiv:1306.1234].




\bibitem{semimetal_effects10}
M. Vazifeh and M. Franz,
{\sl ``Electromagnetic response of weyl semimetals''},
Phys. Rev. Lett. {\bf 111}, 027201 (2013) [arXiv:1303.5784].

\bibitem{semimetal_effects11}
Y. Chen, S. Wu, and A. Burkov,
{\sl ``Axion response in Weyl semimetals''},
Phys. Rev. B {\bf 88}, 125105 (2013) [arXiv:1306.5344].

\bibitem{semimetal_effects12}
Y. Chen, D. Bergman, and A. Burkov,
{\sl ``Weyl fermions and the anomalous Hall effect in metallic ferromagnets''},
Phys. Rev. B {\bf 88}, 125110 (2013) [arXiv:1305.0183];
David Vanderbilt, Ivo Souza, and F. D. M. Haldane
Phys. Rev. B {\bf 89}, 117101 (2014) [arXiv:1312.4200].

\bibitem{semimetal_effects13}
S. T. Ramamurthy and T. L. Hughes,
{\sl ``Patterns of electro-magnetic response in topological semi-metals''},
arXiv:1405.7377.









\bibitem{Zyuzin:2012tv}
A.~A.~Zyuzin and A.~A.~Burkov,
  {\sl ``Topological response in Weyl semimetals and the chiral anomaly,''}  Phys.\ Rev.\ B {\bf 86} (2012) 115133  [arXiv:1206.1868 [cond-mat.mes-hall]].  


\bibitem{tewary}
Pallab Goswami, Sumanta Tewari, {\sl Axionic field theory of (3+1)-dimensional Weyl semi-metals,}
Phys. Rev. B 88, 245107 (2013), arXiv:1210.6352

\bibitem{16}
Chao-Xing Liu, Peng Ye,  Xiao-Liang Qi,
{\sl Chiral gauge field and axial anomaly in a Weyl semimetal},
Physical Review B, vol. 87, Issue 23, id. 235306


 \bibitem{Volovik2003} G.E. Volovik, {\it The Universe in a Helium
Droplet}, Clarendon Press,  Oxford (2003).


\bibitem{Metl}
``Anomalous Axion Interactions and Topological Currents in Dense Matter'',Max A. Metlitski and Ariel R. Zhitnitsky,Phys. Rev. D 72, 045011

\bibitem{CME}
 K. Fukushima, D.E. Kharzeev, H.J. Warringa, Phys.Rev.D 78:074033,2008


\bibitem{Kharzeev:2013ffa}
  D.~E.~Kharzeev,
  ``The Chiral Magnetic Effect and Anomaly-Induced Transport,''
  Prog.\ Part.\ Nucl.\ Phys.\  {\bf 75} (2014) 133
  doi:10.1016/j.ppnp.2014.01.002
  [arXiv:1312.3348 [hep-ph]].



\bibitem{Kharzeev:2015znc}
  D.~E.~Kharzeev, J.~Liao, S.~A.~Voloshin and G.~Wang,
  ``Chiral Magnetic Effect in High-Energy Nuclear Collisions --- A Status Report,''
  arXiv:1511.04050 [hep-ph].

\bibitem{Kharzeev:2009mf}
  D.~E.~Kharzeev,
  ``Chern-Simons current and local parity violation in hot QCD matter,''
  Nucl.\ Phys.\ A {\bf 830} (2009) 543C
  doi:10.1016/j.nuclphysa.2009.10.049
  [arXiv:0908.0314 [hep-ph]].

\bibitem{Vilenkin}  A. Vilenkin, Phys. Rev. D 22, 3080 (1980)

\bibitem{TTKN} D. J. Thouless, M. Kohmoto, M. P. Nightingale, and M. den Nijs,
Phys. Rev. Lett. 49, 405 (1982).


\bibitem{Hall3DTI} X.-L. Qi, T. L. Hughes, and S.-C. Zhang, Physical Review B 78, 195424 (2008).





\bibitem{SonYamamoto2012}
D.T.  Son and N. Yamamoto,
{\sl ''Berry curvature, triangle anomalies, and chiral magnetic effect in Fermi liquids''}, Phys.Rev.Lett.109:181602,2012,
 arXiv:1203.2697.



\bibitem{Kharzeev:2009pj}
  D.~E.~Kharzeev and H.~J.~Warringa,
  ``Chiral Magnetic conductivity,''
  Phys.\ Rev.\ D {\bf 80} (2009) 034028
  doi:10.1103/PhysRevD.80.034028
  [arXiv:0907.5007 [hep-ph]].

\bibitem{ZrTe5}
Q. Li, D. E. Kharzeev, C. Zhang, Y. Huang, I. Pletikosic, A. V. Fedorov, R. D. Zhong, J. A. Schneeloch, G. D. Gu, and T. Valla,
arXiv:1412.6543.



\bibitem{Nielsen:1983rb}
  H.~B.~Nielsen and M.~Ninomiya,
  ``Adler-bell-jackiw Anomaly And Weyl Fermions In Crystal,''
  Phys.\ Lett.\ B {\bf 130} (1983) 389.
  doi:10.1016/0370-2693(83)91529-0

\bibitem{nogo}
 M. M. Vazifeh and M. Franz, Phys. Rev. Lett. 111,
027201 (2013)

\bibitem{nogo2} N. Yamamoto, Phys. Rev. D 92, 085011 (2015).





\bibitem{Z2016_1}
  M.~A.~Zubkov,
{\sl ``Wigner transformation, momentum space topology, and anomalous transport,''}
  Annals Phys.\  {\bf 373}, 298 (2016)
  [arXiv:1603.03665 [cond-mat.mes-hall]].



\bibitem{Z2016_2}
  M.~A.~Zubkov,
{\sl ``Absence of equilibrium chiral magnetic effect,''}
  Phys.\ Rev.\ D {\bf 93}, no. 10, 105036 (2016)
  [arXiv:1605.08724 [hep-ph]].

\bibitem{Wigner}  E.P. Wigner, "On the quantum correction for thermodynamic equilibrium", Phys. Rev. 40 (June 1932) 749–759. doi:10.1103/PhysRev.40.749

\bibitem{star} C. Zachos, D. Fairlie, and T. Curtright, "Quantum Mechanics in Phase Space" ( World Scientific, Singapore, 2005) ISBN 978-981-238-384-6

\bibitem{Weyl}
Robert G Littlejohn, "The semiclassical evolution of wave packets",
Physics Reports Volume 138, Issues 4–5, May 1986, Pages 193-291

\bibitem{berezin} Berezin, F.A. and M.A. Shubin, 1972, in: Colloquia Mathematica Societatis Janos Bolyai (North-Holland, Amsterdam) p. 21.



\bibitem{Volovik:2011kg}
  G.~E.~Volovik,
  ``Topology of quantum vacuum,''
  Lecture Notes in Physics {\bf 870} (2013) 343
  [arXiv:1111.4627 [hep-ph]].


\bibitem{Z2016_3}
  M.~A.~Zubkov,
 {\sl ``Momentum space topology of QCD,''}
 Annals Phys.\  {\bf 393} (2018) 264
  arXiv:1610.08041 [hep-th]





























\bibitem{KZ2017}
  Z.~V.~Khaidukov and M.~A.~Zubkov,
  ``Chiral Separation Effect in lattice regularization,''
  Phys.\ Rev.\ D {\bf 95} (2017) no.7,  074502
  doi:10.1103/PhysRevD.95.074502
  [arXiv:1701.03368 [hep-lat]].

\bibitem{Volovik:2016mre}
  G.~E.~Volovik and M.~A.~Zubkov,
  ``Standard Model as the topological material,''
  New J.\ Phys.\  {\bf 19} (2017) no.1,  015009
  doi:10.1088/1367-2630/aa573d
  [arXiv:1608.07777 [hep-ph]].
	





\bibitem{RotHall} 
  M.~A.~Zubkov,
  ``Hall effect in the presence of rotation,''
  EPL {\bf 121} (2018) no.4,  47001
  doi:10.1209/0295-5075/121/47001
  [arXiv:1801.05368 [hep-ph]].


\bibitem{SME} 
  M.~N.~Chernodub,
  ``Anomalous Transport Due to the Conformal Anomaly,''
  Phys.\ Rev.\ Lett.\  {\bf 117} (2016) no.14,  141601
  doi:10.1103/PhysRevLett.117.141601
  [arXiv:1603.07993 [hep-th]].



\bibitem{CSE}
``Anomalous Axion Interactions and Topological Currents in Dense Matter'',Max A. Metlitski and Ariel R. Zhitnitsky,Phys. Rev. D 72, 045011


\bibitem{vlad}D. Diakonov, A.G. Tumanov, A.A. Vladimirov, Low-energy General Relativity with torsion: a systematic
derivative expansion, Phys. Rev. D 84, 124042 (2011); D. Diakonov, Towards lattice-regularized
quantum gravity, arXiv:1109.0091.

\bibitem{VolZub} G.E.Volovik, M.A.Zubkov,	Annals of Physics 340/1 (2014), pp. 352-368,	arXiv:1305.4665 [cond-mat.mes-hall].

\bibitem{Voz} F. de Juan, A. Cortijo and M.A.H. Vozmediano, Dislocations and torsion in graphene and related
systems, Nucl. Phys. B 828, 625–637 (2010).

\bibitem{Wigner}Wigner transformation, momentum space topology, and anomalous transport
M.A Zubkov.Annals of Physics.Volume 373, Pages 298–324

\bibitem{CSEwith}Z.~V.~Khaidukov and M.~A.~Zubkov,
  ``Chiral Separation Effect in lattice regularization,''
  Phys.\ Rev.\ D {\bf 95} (2017) no.7,  074502
  doi:10.1103/PhysRevD.95.074502
  [arXiv:1701.03368 [hep-lat]].

\bibitem{ACVE} Ruslan Abramchuk, Z.V.Khaidukov, M.A. Zubkov, arXiv:1806.02605 [hep-ph]


\bibitem{Land}K.~Landsteiner, E.~Megias and F.~Pena-Benitez,
  ``Gravitational Anomaly and Transport,''
  Phys.\ Rev.\ Lett.\  {\bf 107} (2011) 021601
  doi:10.1103/PhysRevLett.107.021601
  [arXiv:1103.5006 [hep-ph]].

\bibitem{Sumiyoshi:2015eda}
  H.~Sumiyoshi and S.~Fujimoto,
  ``Torsional Chiral Magnetic Effect in a Weyl Semimetal with a Topological Defect,''
  Phys.\ Rev.\ Lett.\  {\bf 116} (2016) no.16,  166601
  doi:10.1103/PhysRevLett.116.166601
  [arXiv:1509.03981 [cond-mat.mes-hall]].


\bibitem{Nieh} H. T. Nieh and M. L. Yan, Ann. Phys. 138, 237 (1982).
H. T. Nieh, A torsional topological invariant, International Journal of Modern Physics A 22, 5237–5244
(2007)

\bibitem{Parrikar}
Onkar Parrikar, Taylor L. Hughes, and Robert G. Leigh, "Torsion, parity-odd response,
and anomalies in topological states," Phys. Rev. D 90, 105004 (2014)

\bibitem{Cortijo:2015jja} A.~Cortijo and M.~A.~Zubkov, \textit{``Emergent
gravity in the cubic tight-binding model of Weyl semimetal in the presence
of elastic deformations,''} Annals Phys.\ \textbf{366} (2016) 45
doi:10.1016/j.aop.2016.01.006 [arXiv:1508.04462 [cond-mat.mes-hall]].

\end{thebibliography}
\end{document}